\documentclass[aps, pra,superscriptaddress,longbibliography]{revtex4-1}%
\usepackage{graphicx}
\usepackage{float}
\usepackage{dcolumn}
\usepackage{bm,color}
\usepackage{amsmath,amssymb,dsfont,amstext,amsfonts}
\usepackage{extarrows}
\usepackage{gensymb}
\usepackage[colorlinks=true,linkcolor=blue,urlcolor=blue,citecolor=blue]{hyperref}
\usepackage{xcolor}
\usepackage{wasysym}
\usepackage{mathtools}
\usepackage{bbold}
\usepackage{tikz}

\usepackage{graphicx}
\graphicspath{ {./images/} }

\usepackage[normalem]{ulem} 
\usepackage{color}

\usepackage[labelsep=endash]{caption}

\begin{document}
\title{{Helicity-dependent Ultrafast Photocurrents in Weyl Magnet Mn$_3$Sn}}
\author{Dominik Hamara$^{\S}$}
%\email{dh541@cam.ac.uk}
\affiliation{Cavendish Laboratory, University of Cambridge, J.J.~Thomson Avenue, Cambridge CB3 0HE, United Kingdom}
\author{Gunnar F. Lange$^{\S}$}
%\email{gfl25@cam.ac.uk}
\affiliation{Cavendish Laboratory, University of Cambridge, J.J.~Thomson Avenue, Cambridge CB3 0HE, United Kingdom}
\author{Farhan Nur Kholid}
\affiliation{Cavendish Laboratory, University of Cambridge, J.J.~Thomson Avenue, Cambridge CB3 0HE, United Kingdom}
\author{Anastasios Markou}
%\email{anastasios.markou@cpfs.mpg.de} 
\affiliation{Max Planck Institute for Chemical Physics of Solids, 01187 Dresden, Germany}
\affiliation{Physics Department, University of Ioannina, 45110 Ioannina, Greece}
\author{Claudia Felser}
%\email{anastasios.markou@cpfs.mpg.de} 
\affiliation{Max Planck Institute for Chemical Physics of Solids, 01187 Dresden, Germany}
\author{Robert-Jan Slager}
\email{rjs269@cam.ac.uk} 
\affiliation{Cavendish Laboratory, University of Cambridge, J.J.~Thomson Avenue, Cambridge CB3 0HE, United Kingdom}
\author{Chiara Ciccarelli}
\email{cc538@cam.ac.uk}
\affiliation{Cavendish Laboratory, University of Cambridge, J.J.~Thomson Avenue, Cambridge CB3 0HE, United Kingdom}

%%TC:ignore
\date{\today}
\begin{abstract}
$^{\S}$ These authors contributed equally.
\\
\\
We present an optical pump-THz emission study on non-collinear antiferromagnet Mn$_3$Sn. We show that Mn$_3$Sn acts as a source of THz radiation when irradiated by femtosecond laser pulses. The polarity and amplitude of the emitted THz fields can be fully controlled by the polarisation of optical excitation. We explain the THz emission with the photocurrents generated via the photon drag effect by combining various experimental measurements as a function of pump polarisation, magnetic field, and sample orientation with thorough symmetry analysis of response tensors.

\end{abstract}

\maketitle
\section{Introduction}\label{sec:introduction}
Mn$_3$Sn is a noncollinear antiferromagnet (AF) and a Weyl semimetal (WSM). It crystallises in a hexagonal P6$_3$/mmc structure. Below the Néel temperature ($T_N$ $\approx$ 420 K for bulk Mn$_3$Sn \cite{Sung2018}) the geometrical frustration of the Mn atoms in the $a$-$b$ plane of the Kagome lattice leads to an inverse triangular spin structure, with 120° ordering \cite{Brown1990,Sung2018, Yang2017, Li2018, Cheng2019}. Despite a vanishingly small net magnetisation, Mn$_3$Sn displays phenomena that conventionally occur in ferromagnets, such as a large anomalous Hall effect \cite{Kubler_magnetism_2014, Nakatsuji2015, Nayak2016}, anomalous Nernst effect \cite{Ikhlas2017}, and magneto-optical Kerr effect \cite{Higo2018}. This is possible due to the unique material topology and a nonzero Berry curvature resulting from the inverse triangular spin structure \cite{Sung2018, Nayak2016}. Ab-initio band structure calculations have reported the existence of multiple Weyl points in the bulk and corresponding Fermi arcs on the surface of Mn$_3$Sn \cite{Yang2017}. An effect associated with WSMs is the presence of helicity-dependent photocurrents arising from non-linear optical effects. These have been observed using both electrical and THz techniques \cite{Chan2017}, and have been linked to the topological charge of the Weyl nodes \cite{deJuan_2017}, via the circular photogalvanic effect.
\par
In this work, we present the experimental observation of helicity-dependent ultrafast photocurrents in an 80 nm Mn$_3$Sn film at room temperature (RT) using optical pump-THz emission spectroscopy. The magnitude and direction of the photocurrents depend on the polarisation of the pump pulse and the direction of its wavevector relative to the surface of the film, but have no dependence on magnetic field. These currents cannot be attributed to a bulk photogalvanic effect as this requires the breaking of inversion symmetry \cite{Le2021}. Mn$_3$Sn, however, respects inversion symmetry even when accounting for the magnetic ordering. This suggests that our signal originates either from a different bulk mechanism such as the inverse spin Hall effect \cite{Hirai2020} or photon-drag effect \cite{Ribakovs1977,Maysonnave2014}, or from a surface photogalvanic effect \cite{Steiner2022}. Our symmetry analysis of response tensors suggests that the helicity-dependent photocurrents arise predominantly due to the circular photon drag effect. 
\par

\section{Experimental Results}
\subsection{The sample}

\par
The subject of the study is an MgO(111)(0.5 mm)/Ru(5 nm)/Mn$_3$Sn(80 nm)/Si(3 nm) sample. Epitaxial Mn$_3$Sn films were grown using magnetron sputtering in a BESTEC ultra-high vacuum (UHV) system with a base pressure less than $2\times 10^{-9}$\,mbar and a process gas (Ar 5 N) pressure of $3 \times 10^{-3}$\,mbar. The target to substrate distance was fixed at $20$\,cm and the substrates were rotated during deposition to ensure homogeneous growth. The underlayer was deposited using a Ru (5.08\,cm) target by applying 40\,W DC power with the substrate held at 400$^{\circ}$\,C. Following cooling back to room temperature, Mn$_3$Sn was grown from Mn (7.62\,cm) and Sn (5.08\,cm) sources in confocal geometry, using 47\,W and 11\,W DC power respectively. The stack was then annealed in-situ under UHV at 350$^{\circ}$\,C for 10 minutes. The stoichiometry is Mn$_{75}$Sn$_{25}$ (± 2 at. \%), estimated by using energy dispersive x-ray spectroscopy (see SI). Finally, a Si capping layer was deposited at room temperature using an Si (5.08\,cm) target at 60\,W RF power to protect the film from oxidation. Magnetotransport studies on films grown under the same conditions and with comparable crystal quality are presented in \cite{Taylor2020} and show a large anomalous Hall effect at room temperature
and a transition to topological Hall effect below 50\,K.

\subsection{Experimental layout}

In Fig.~\ref{fig:fig1}(a) we show the optical pump-THz emission geometries used in our experiments. When presenting a data set we will refer to the experiment geometry in which this data was collected as \textit{configuration} \textit{1}, \textit{2} or \textit{3}. To indicate different directions we introduce two Cartesian coordinate systems, also indicated in Fig.~\ref{fig:fig1}(a): ($x$, $y$, $z$) - fixed with respect to the experimental setup; ($a$, $b$, $c$) - fixed with respect to the sample. Laser pulses of 50\,fs duration with a central wavelength of 800\,nm propagate along the $z$ axis. The optical fluence is fixed at 2.9\,mJ/cm$^2$. A quarter-wave plate (QWP) placed in the pump path is used to control the polarisation and helicity of the pulses. In-plane rotations of the QWP by an angle $\varphi$ allow changing between linear ($\varphi=0\degree \pm n\pi/2$), left-handed circular (LHCP) ($\varphi=45\degree \pm n\pi$), right-handed circular (RHCP) ($\varphi=-45\degree \pm n\pi$), and intermediate elliptical polarisations. The angle between the laser beam and the sample surface can be varied by rotating the sample away from normal incidence about the $x$ axis (\textit{configuration 2}) or $y$ axis (\textit{configuration 3}). We define the tilting angles as $\theta_{x}$ and $\theta_{y}$ respectively. Additionally, the sample can be rotated in-plane, about the $c$ axis by an angle defined as $\theta_{c}$. An external magnetic field up to $\mu_0 H_{x}\approx $ 860\,mT can be applied along the $x$ direction. Unless stated otherwise, the experiments presented in the main body of the paper were performed at room temperature (RT).

\subsection{THz emission from Mn$_3$Sn}

The optically induced charge currents $J_i(t)$ result in broadband THz electro-dipole emission $s_i(t)$. The subscript $i$ indicates the polarisation components $i=x$ or $i=y$, along the two axes of the experimental setup $x$ and $y$. $S_i$ is defined as the integrated peak amplitude of the emitted THz pulse.
\par
\begin{figure}[h]
    \centering
    \includegraphics[width=1\textwidth]{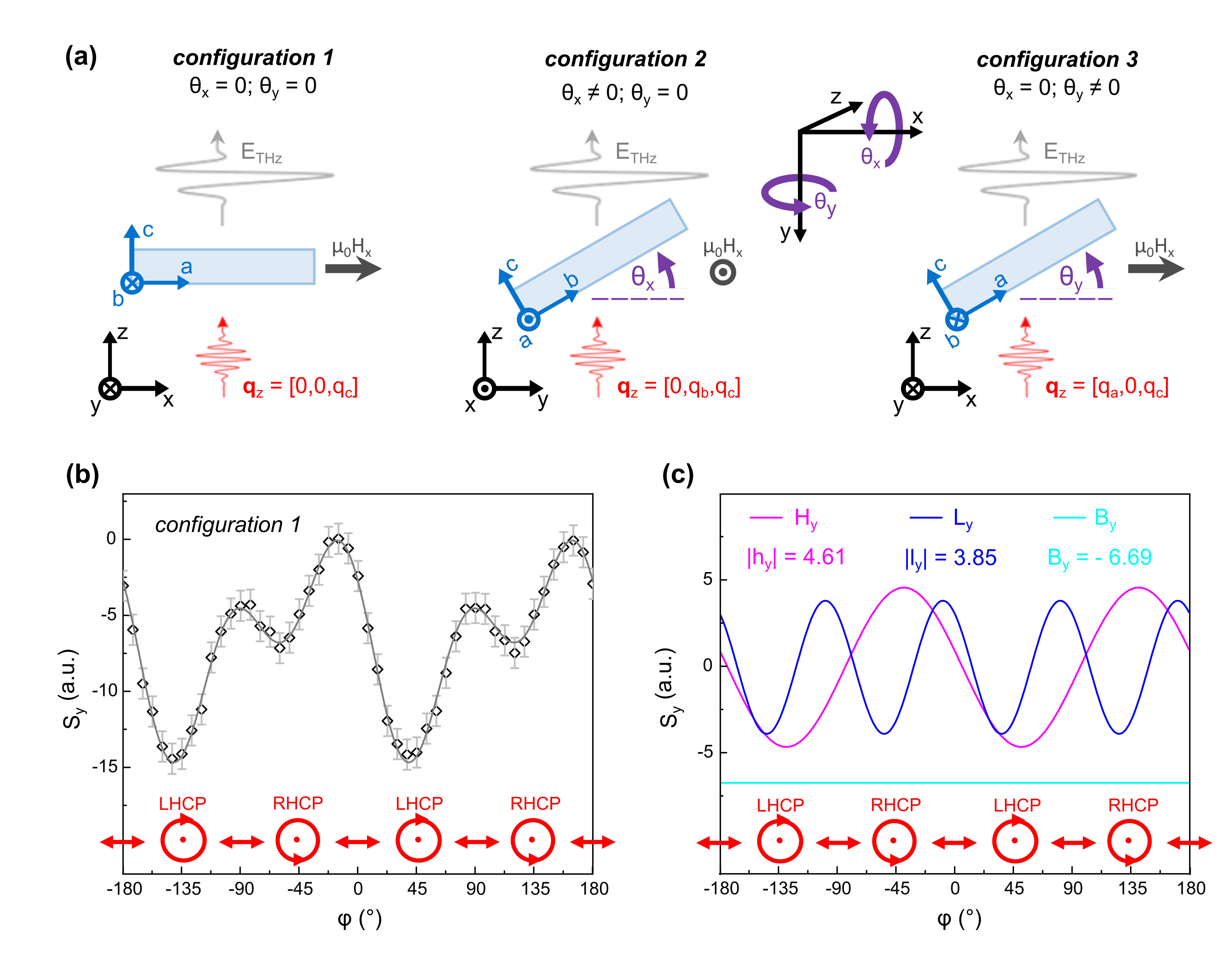}
    \caption{Optical pump-THz emission from Mn$_3$Sn. \textbf{(a)} Experiment schemes in different geometries. Ultrafast laser pulses travelling along $z$ axis are incident on the sample, leading to the emission of $E_i(t)$ fields. In experiments, we resolve their two orthogonal components of transient THz signals: $s_{x}(t) \propto E_{x}(t)$ and $s_{y}$(t)$\propto E_{y}(t)$. \textit{Configuration 1} corresponds to a measurement in the standard geometry with normal laser incidence, in which $\theta_{x}=0$ and $\theta_{y}=0$. In \textit{configuration 2} and \textit{configuration 3} the sample is rotated about the $x$ ($\theta_{x}\neq0$ and $\theta_{y}=0$) or $y$ ($\theta_{x}=0$ and $\theta_{y}\neq0$) axes respectively. Please note that the scheme of \textit{configuration 2} is presented in a different perspective than \textit{configuration 1} and \textit{3}. \textbf{(b)} shows THz signals detected along $y$, emitted from Mn$_3$Sn optically pumped with laser pulses of different polarisations. The measurements were performed in \textit{configuration 1} without an external magnetic field. The experimental data set is fitted with function introduced in Eq. \ref{eqn:Eq2} to find $H_{y}$, $L_{y}$, and $B_{y}$. These are presented in \textbf{(c)} with the extracted values of $|h_{y}|$, $|l_{y}|$, and $B_{y}$.}
    \label{fig:fig1}
\end{figure}

Fig. \ref{fig:fig1}(b) shows the $\varphi$-dependence of $S_{y}$, measured at zero magnetic field ($\mu_0 H_{x}=0$) and at normal pump incidence (\textit{configuration 1}). The different polarisations of the pump pulse that correspond to the different values of $\varphi$ are also indicated to facilitate the reading. The data is decomposed into different harmonics by fitting with the equation \cite{Ji2019}:
\begin{equation}
\label{eqn:Eq2}
    S_i(\varphi) = H_i(\varphi)+L_i(\varphi)+B_i
\end{equation}

Here, $H_i(\varphi)=h_i \sin(2\varphi + \varphi_1)$ is the magnitude of the circular polarisation helicity-dependent component with a phase shift $\varphi_1$, $L_i(\varphi)=l_i \sin(4\varphi + \varphi_2)$ is the linear polarisation-dependent component with a phase shift $\varphi_2$, and $B_i$ is the polarisation-independent background component. As displayed in Fig.~\ref{fig:fig1}(c), the decomposition of $S_{y}$ shows that $H_{y}(\varphi)$, $L_{y}(\varphi)$, and $B_{y}$ all contribute to the measured signal. For comparison, $S_{x}$ measured in the same experimental geometry (\textit{configuration 1}) displays a relatively smaller contribution from the helicity-dependent $H_{x}(\varphi)$ component, and is dominated by $L_{x}(\varphi)$ (Fig.~\ref{fig:figS1} in Supplementary Information). We attribute this difference to a small unintentional rotation around the $y$ axis, as it will be further justified in the analysis that follows. In our set-up, the sample's mount orientation can be freely adjusted around the $y$ axis, but not around the $x$ axis, so an unintentional tilting around the $y$ axis is more likely. 

\subsection{The effect of experiment geometry}

Here we study how both polarisation and amplitude of the photocurrent-generated THz emission depend on the direction of the optical wavevector $q_z$ relative to the sample surface. For this purpose we tilt the sample whilst leaving the direction of the pump wavevector  $q_{z}$ unchanged with respect to the laboratory frame of reference, along the $z$-direction. For $\theta_{x}$ (\textit{configuration 2}) or $\theta_{y}$ (\textit{configuration 3}) different from zero, $q_{z}$ has a non-zero projection along the $b$ and $a$ directions on the plane of the Mn$_3$Sn film, which we label $q_b$ and $q_a$ respectively. Consequently, the components of the wavevector relative to the sample frame are $[0,q_b,q_c]$ in \textit{configuration 2} and and $[q_a,0,q_c]$ in \textit{configuration 3}.

\par
Fig.~\ref{fig:fig2}(a) and (b) show the components of the THz emission polarised along the $x$ and the $y$ directions as a function of $\varphi$, measured in \textit{configuration 2} (a) and \textit{configuration 3} (b). $S_{x}(\varphi)$ and $S_{y}(\varphi)$ are fitted with Eq.~\eqref{eqn:Eq2} to extract the coefficients $h_{x}$ and $h_{y}$. While in Fig.~\ref{fig:fig2}(a) $h_{x}\gg h_{y}$, in Fig.~\ref{fig:fig2}(b) the trend is inverted and  $h_{x}\ll h_{y}$. This suggests that the direction of the helicity-dependent photocurrent, and therefore of the THz polarisation, depends on the projection of the pump wavevector on the sample plane and is perpendicular to it.
Due to finer control of $\theta_{y}$ in comparison to $\theta_{x}$ in our setup, we restrict the following analysis to \textit{configuration 3} only. In Fig.~\ref{fig:fig2}(c) we show that the THz emission amplitude increases with tilting angle $\theta_y$.
We now study the symmetry of the THz emission for opposite tilting angles by decomposing $S_{i}(\varphi)$ into even and odd contributions as:

\begin{equation}
\label{eqn:Eq3}
    \begin{split}
S_i(\varphi)_{even}(\theta_{y} = \pm 15 \degree) 
        = S_i(\varphi)(\theta_{y} = +15 \degree)
        + S_i(\varphi)(\theta_{y} = -15 \degree)
   \end{split}
\end{equation}

\begin{equation}
\label{eqn:Eq4}
    \begin{split}
S_i(\varphi)_\text{odd}(\theta_{y} = \pm 15 \degree) = S_i(\varphi)(\theta_{y} = +15 \degree)-S_i(\varphi)(\theta_{y} = -15 \degree)
   \end{split}
\end{equation}

 In Fig.~\ref{fig:fig2}(d) We observe that the odd component of $S_y$ is dominant. Analogous behaviour is presented in the Supplementary Information (SI) for $S_x$. Our observations suggest that the helicity-dependent photocurrents are generated in the direction perpendicular to the in-plane projection of the pump wavevector and are proportional to it.
\par
\subsection{Magnetic field dependence}

In this section we want to understand whether the photocurrents have a magnetic origin and depend on the magnetic phase of Mn$_3$Sn. Fig.~\ref{fig:fig3}(a) shows $S_{y}$ measured for two opposite directions of the magnetic field $\pm 0.860$\,mT. According to Reichlova \textit{et al.} \cite{Reichlova2019}, $0.860$\,mT  may be too low to switch the magnetic ordering in Mn$_3$Sn thin films at RT, while it is sufficient to reverse the spins at temperatures close to $T_N$. Hence, for these measurements we followed a cool down procedure with the magnetic field continuously applied from 420 K to 380 K, at which the experiment was performed. No significant dependence on magnetic field is measured. We also do not observe a qualitative difference in the time-domain THz transients measured at different fields as shown in the SI.

We further confirm that the polarisation and amplitude of the emitted THz pulse is not correlated with the magnetic phase of Mn$_3$Sn by repeating the measurement after rotating the sample by $90\degree$ around the $c$ axis. 
Fig.~\ref{fig:fig3}(b) shows $S_{x}(\varphi)$ and $S_{y}(\varphi)$ prior and after the rotation by $\theta_c = +90 \degree$.
If the direction of the photoinduced currents, hence the polarisation of the THz emission, were correlated with the orientation of the spins we would have expected a rotation of the THz polarisation plane by $90\degree$, which we do not observe. Instead, the two graphs of $S_{x}(\varphi)$ and $S_{y}(\varphi)$ overlap, as is discussed further in the SI. 

\section{Theoretical analysis}
\subsection{Nonlinear optical effects}
In this section, we investigate nonlinear optical effects in Mn$_3$Sn as an explanation of the observed signal. Other possible sources of helicity-dependent photocurrents are discussed in the subsequent section.
\par
We consider a phenomenological expression for the induced photocurrent $J_i$ that in turn generates THz emission via electro-dipole interaction. Using the notation from Ref.~\cite{Hamh2016} and expanding to second order in the light wave amplitude $E(\omega)$ for a frequency $\omega$ we write:
\begin{equation}\label{eq:response_tensor_general_eq}
    J_i = \chi_{ijk}^{(2)}E_jE_k^* + \chi_{ijkl}^{(3)}q_jE_kE_l^*
\end{equation}
where all indices run over the Cartesian coordinates of the sample $i,j,k,l\in (a,b,c)$ and $\boldsymbol{q}$ is the momentum of the incoming light.  At normal incidence (\textit{configuration 1}), only $q_c, E_a$ and $E_b$ are non-zero. The first term describes the photogalvanic effect (PGE), whereas the second term describes the photon-drag effect (PDE). We further decompose each of these tensors into symmetric and antisymmetric components with respect to the light wave amplitude which respectively give rise to the linear photogalvanic/photon drag effect (LPGE/LPDE) and the circular photogalvanic/photon drag effect (CPGE/CPDE) \cite{Karch2010}. Note that all quantitative features of these effects depend crucially on the details of the band structure. We show the ab-initio bulk and surface band structures in Fig.~\ref{fig:fig4}, for an energy-window of $\hbar \omega \approx 1.55$\,eV, corresponding to the central wavelength of the laser pulses. Due to the large number of bands involved, we focus on a qualitative phenomenological symmetry analysis of the tensors in Eq.~\eqref{eq:response_tensor_general_eq}.

\subsubsection{Symmetry-constrained model}
The spatial symmetries of the material constrain the tensors in Eq.~\eqref{eq:response_tensor_general_eq} and topology~\cite{Kruthoff2019,Schrunk}. Mn$_3$Sn has space-group symmetry P$6_3/$mmc when ignoring magnetism and magnetic space-group symmetry C$m'$c$m'$ when including the AFM ordering \cite{Brown1990}. For the tensor symmetry analysis, we focus on the unitary point-group symmetries as detailed in the SI, where also complete expressions for the symmetry-allowed forms of the PGE/PDE tensors in Eq.~\eqref{eq:response_tensor_general_eq} are given. Here, we only summarise the number of independent coefficients for the various symmetry settings as shown in Tab.~\ref{tab:Symm_coeff}. In general, there are too many possible terms, making a quantitative model infeasible. However, by carefully comparing with our experimental results, it is possible to identify effects are the most relevant as detailed in the next section and in the SI.
\begin{table}[ht!]
    \centering
    \begin{tabular}{c c |c c c c}
    PG & Relevance & LPGE & CPGE & LPDE & CPDE\\
      6/mmm & NM bulk & 0 & 0  &7 & 3 \\
    3m & NM surface & 4 & 1 & 10 & 4\\ 
    \hline
    2/m &M bulk& 0 & 0 & 28 &13\\
    1 &M surface & 18 & 9 & 54 & 27\\ 
        
    \end{tabular}
    \caption{Number of independent elements for the linear/circular photogalvanic effect (LPGE/CPGE) and the linear/circular photon drag effect (LPDE/CPDE), for various unitary point-group (PG) symmetries relevant to the non-magnetic/magnetic (NM/M) bulk/surface of the material.}
    \label{tab:Symm_coeff}
\end{table}

\subsubsection{Interpretation of results}
We begin by considering the effect of magnetism. As shown in Fig.~\ref{fig:fig3}(a) our results are insensitive to the direction and magnitude of the external magnetic field, as well as to the intrinsic spin ordering of the material [see Fig.~\ref{fig:fig3}(b)]. This suggests that the generated photocurrents do not arise as a result of the magnetic ordering in the material. We therefore focus on the non-magnetic symmetry analysis in what follows. Because the bulk of the sample respects inversion symmetry we should not expect any bulk contribution from the PGE in this case.

We address here the result of tilting the sample away from the normal pump incidence, as shown in Fig.~\ref{fig:fig2}. As discussed in detail in the SI, the tilting changes both what currents are generated in the material, and which part of the resultant THz radiation is measured at the detector. Tilting in $\theta_x$ (\textit{configuration 2}) or $\theta_y$ (\textit{configuration 3}) changes the geometry of the sample relative to the detector. In particular, we are able to resolve currents generated in the $c$-direction \cite{Ni2021}. Thus, the detected integrated amplitude of the THz transient pulse in \textit{configuration 2} is given by:
\begin{equation}
    \begin{split}
    \begin{gathered}
    S_x \propto J_a(\theta_x)\\
    S_y \propto J_b(\theta_x) \cos \theta_x + J_c(\theta_x) \sin \theta_x
    \end{gathered}
    \end{split}
\end{equation}
And in \textit{configuration 3}:
\begin{equation}
    \begin{split}
    \begin{gathered}
    S_x \propto J_a(\theta_y)\cos \theta_y + J_c (\theta_y)\sin \theta_y\\
    S_y \propto J_b (\theta_y)
    \end{gathered}
    \end{split}
\end{equation}
Where the expressions for $J_i(\theta_k)$ arising from PGE/PDE are give in the SI. When considering the contribution from the bulk photon drag effect and the surface photogalvanic effect to $\boldsymbol{J}$, we find that all terms in the detected amplitude $S_i$ arising from a bulk photon-drag effect are odd under tilting in both $\theta_x$ and $\theta_y$, whereas the linear surface photogalvanic effect contains lowest-order terms that are even under tilting.
As the odd components make a significant contribution to our results [see e.g. Fig.~\ref{fig:fig1}(c)], we interpret our signal to arise predominantly from the bulk photon drag effect.  As discussed in the SI, we further find that the circular photon drag contribution to $S_i$ normal to the rotation axis is suppressed, in agreement with the experimental results, and that only the bulk non-magnetic photon drag contribution is invariant under in-plane rotation [see Fig.~\ref{fig:fig3}(b)]. We therefore interpret our signal to predominantly arise from a bulk photon drag effect.

We note finally that the photon drag effect is also appealing from a bulk band-structure perspective. As shown in Fig.~\ref{fig:fig4}(b), Mn$_3$Sn has a flat bulk band below the Fermi energy around the K-point, with a corresponding band at a distance of $\hbar \omega$. This may lead to a large joint density of states, and allow for non-vertical transitions with finite $\boldsymbol{q}$. 

\subsection{Other possible sources of photocurrent}
Another way in which helicity-dependent photocurrents can be generated is through a combination of the inverse Faraday effect (IFE) and inverse spin Hall effect (ISHE), as described recently for Bismuth thin-films in Ref.~\cite{Hirai2020}. This could be an important mechanism in Mn$_3$Sn, where effects related to the nonzero Berry curvature are significant \cite{Nayak2016} and may result in strong responses of the Faraday effect \cite{Yang2014}. Furthermore, Mn$_3$Sn has been shown to have a large spin Hall angle \cite{Matsuda2020}. However, we believe this mechanism does not play a significant role in our experimental results. Firstly, for an efficient conversion of IFE-induced spin currents into charge currents and THz electric fields the spins must travel relatively long distances \cite{Hirai2020}. This is not the case in Mn$_3$Sn, where the reported spin propagation length is below 1\,nm \cite{Muduli2019}. Secondly, a Berry curvature related effect, such as the IFE, would show a strong dependence on the magnetic phase of Mn$_3$Sn. We do not observe this behaviour in our temperature-resolved measurements (see Fig.~\ref{fig:figS4} in the SI). Previous studies have reported that at temperatures above 420 K Mn$_3$Sn becomes paramagnetic and that upon cooling below RT the material can undergo transitions into the spiral and spin glass phases \cite{Sung2018}. Our investigation in the temperature range of $50-400$ K does not reveal any abrupt changes in the magnitude of THz signals that could indicate the role of magnetic phase-dependent mechanisms. We do not, however, rule out that our sample remains in the same magnetic state over the entire investigated temperature range. Finally, the mechanism relying on the IFE and ISHE could only explain the helicity-dependent photocurrents and cannot account for the generation of photocurrents that show linear dependence on the pump polarisation. As shown in Fig.~\ref{fig:fig1}(c), the magnitudes of $H_y$ and $L_y$ are comparable, and therefore we suggest their main contributions originate from related mechanisms, namely the CPDE and LPDE.

\section{Summary}
In conclusion, using optical pump-THz emission spectroscopy we demonstrate the generation of helicity-dependent ultrafast photocurrents in a Mn$_3$Sn thin film. The magnitude and direction of these can be fully controlled by the polarisation and incidence angle of the optical pump and are not affected by external magnetic fields. We combine the experimental results with theoretical analysis to suggest that the bulk photon drag effect is the main mechanism responsible for the generation of the helicity-dependent photocurrents.

\section*{Methods}
    The electronic band structure was calculated using density-functional theory (DFT) as implemented in Quantum Espresso \cite{Giannozzi2009,Giannozzi2017} with a fully-relativistic norm-conserving pseudopotential, generated using the \texttt{ONCVPSP} package \cite{Hamann2013}. We used the experimental crystal parameters  $a=b=5.67$\,Å and $c = 4.53$\,Å, with an $8\times 8 \times 8$ \textbf{k}-grid and a kinetic-energy cutoff of 870\,eV. The magnetic structure was relaxed by constraining the total direction of the magnetization. The bands were then Wannierised using Wannier90 \cite{Mostofi2014}, with all d-orbitals of Mn considered in the projector. Finally, the slab band structure was computed using WannierTools \cite{WU2017}.

\begin{acknowledgements}
C.~C. and D.~H. acknowledge support from the Royal Society. G.~F.~L is funded by the Aker Scholarship. R.~J.~S acknowledges funding from a New Investigator Award, EPSRC grant EP/W00187X/1, as well as Trinity college, Cambridge. G.~F.~L thanks B. Peng, S. Chen, U. Haeusler and M.A.S. Martínez for useful discussions.
\end{acknowledgements}

\par

\newpage
\maketitle
\section{figures}\label{sec:figures}

\begin{figure}[h]
    \centering
    \includegraphics[width=1\textwidth]{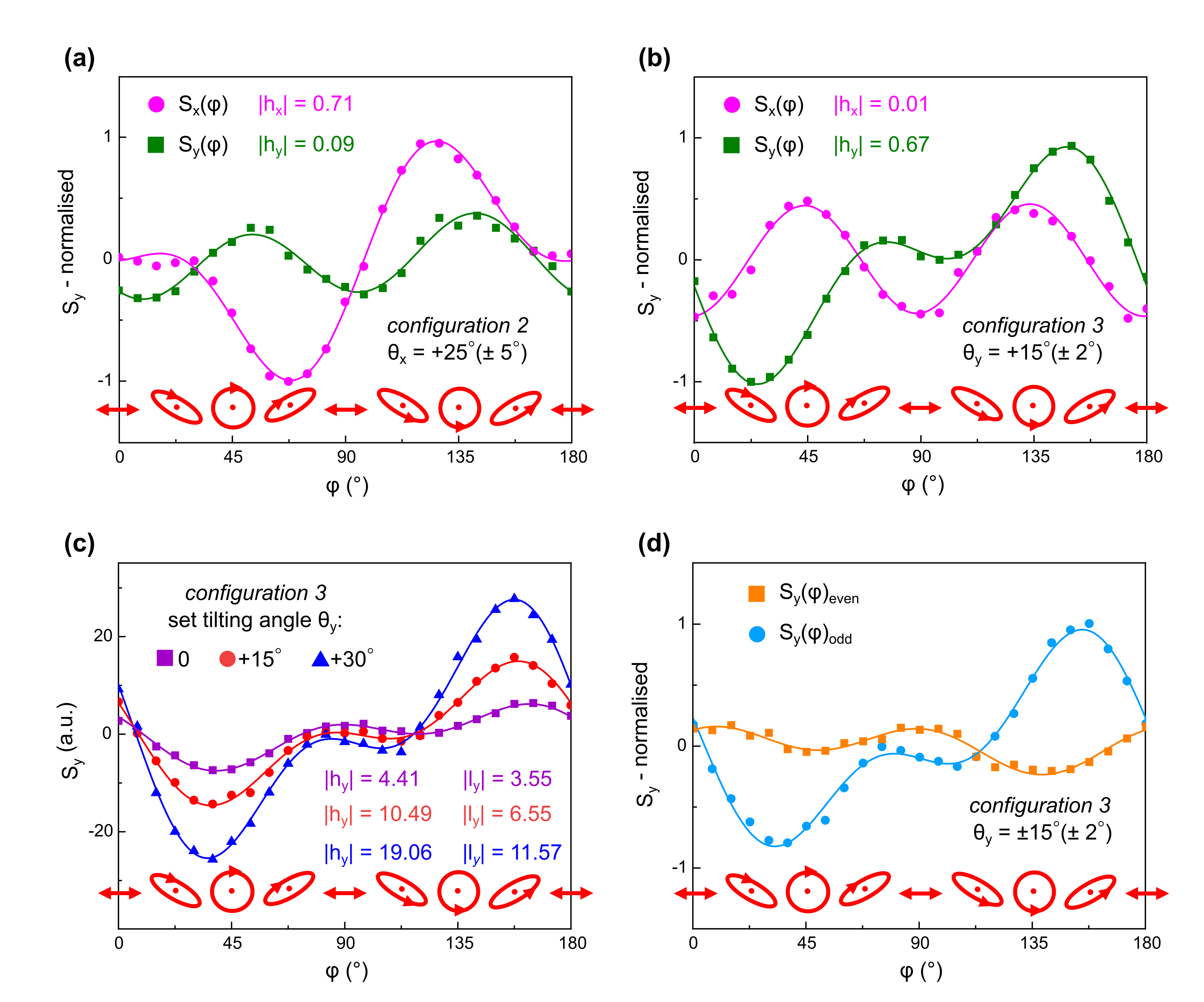}
    \caption{Study of various experiment geometries in optical pump-THz emission from Mn$_3$Sn. Normalised $S_{x}(\varphi)$ and $S_{y}(\varphi)$ data sets measured in \textit{configuration 2} and \textit{configuration 3} are plotted in \textbf{(a)} and \textbf{(b)} respectively. The sample was rotated by $\theta_{x}=+25\degree \pm 5\degree$ or $\theta_{y}=+15\degree \pm 2\degree$ respectively. \textbf{(c)} shows THz signals measured along $y$ in \textit{configuration 3} for tilting angles $\theta_{y}$ between 0 and 30\degree. The data sets were fitted with Eq. \ref{eqn:Eq2} to extract the values of $|h_{y}|$ and $|l_{y}|$ parameters. Extracted $|h_{x}|$ and $|h_{y}|$ are displayed in the graphs. \textbf{(d)} shows even (see Eq. \ref{eqn:Eq3}) and odd (see Eq. \ref{eqn:Eq4}) responses in respect to the direction of the rotation around the $\theta_{y}$ axis. The figure displays normalised THz signals measured along $y$ in \textit {configuration 3}. All measurements in (\textbf{a}-\textbf{d}) were performed at RT with no magnetic field. Data sets shown in \textbf{(a)}, \textbf{(b)} and \textbf{(d)} were fitted with Eq. \ref{eqn:Eq2} prior to normalisation to subtract the polarisation-independent backgrounds, $B_{i}$.}     
    \label{fig:fig2}
\end{figure}

\begin{figure}[h]
    \centering
    \includegraphics[width=1\textwidth]{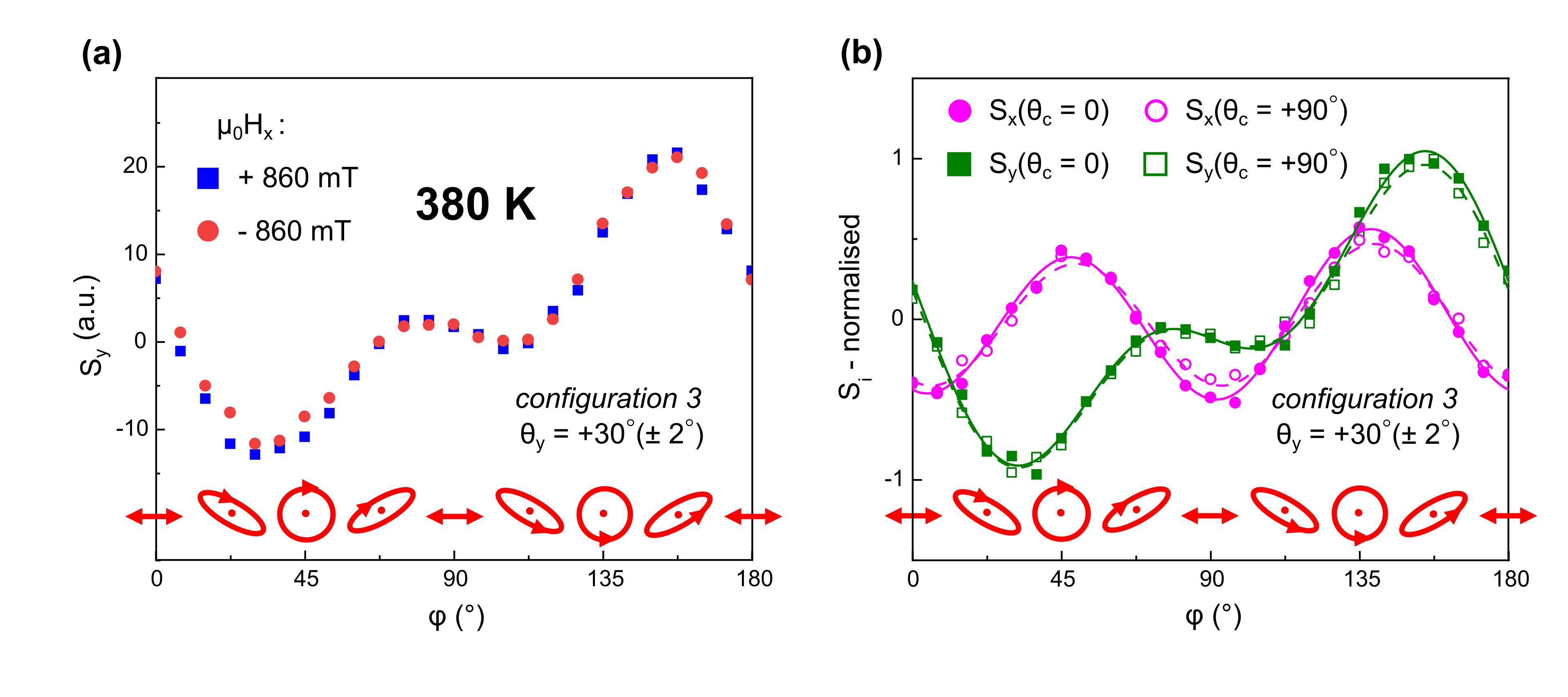}
    \caption{Magnetic field dependence of THz emission from Mn$_3$Sn. \textbf{(a)} 
    THz signals measured at 380 K in \textit{configuration 3} after field-cooling from 420 K with opposite directions of external magnetic field continuously applied along $x$. \textbf{(b)} Normalised $S_{x}(\varphi)$ and $S_{y}(\varphi)$ data sets obtained prior, and after an in-plane rotation of the film by $90\degree$. Prior to the measurements the sample was cooled down from 420 K to RT with $\mu_0 H_{x}=+860\text{mT}$ applied. The field was switched off at RT, and $S_{i}(\varphi)(\theta_{c}=0)$ were measured. Without repeating the field-cooling procedure and at $\mu_0 H_{x}=0$, the film was rotated i-p by $+90\degree$ to obtain $S_{i}(\varphi)(\theta_{c}=+90\degree)$. The experiment was performed in \textit{configuration 3}. The plots show normalised signal values. Prior to normalisation the data sets were fitted with Eq. \ref{eqn:Eq2} to subtract the polarisation-independent backgrounds, $B_{i}$.}
    \label{fig:fig3}
\end{figure}

\begin{figure}[h]
    \centering
    \includegraphics[width=1\textwidth]{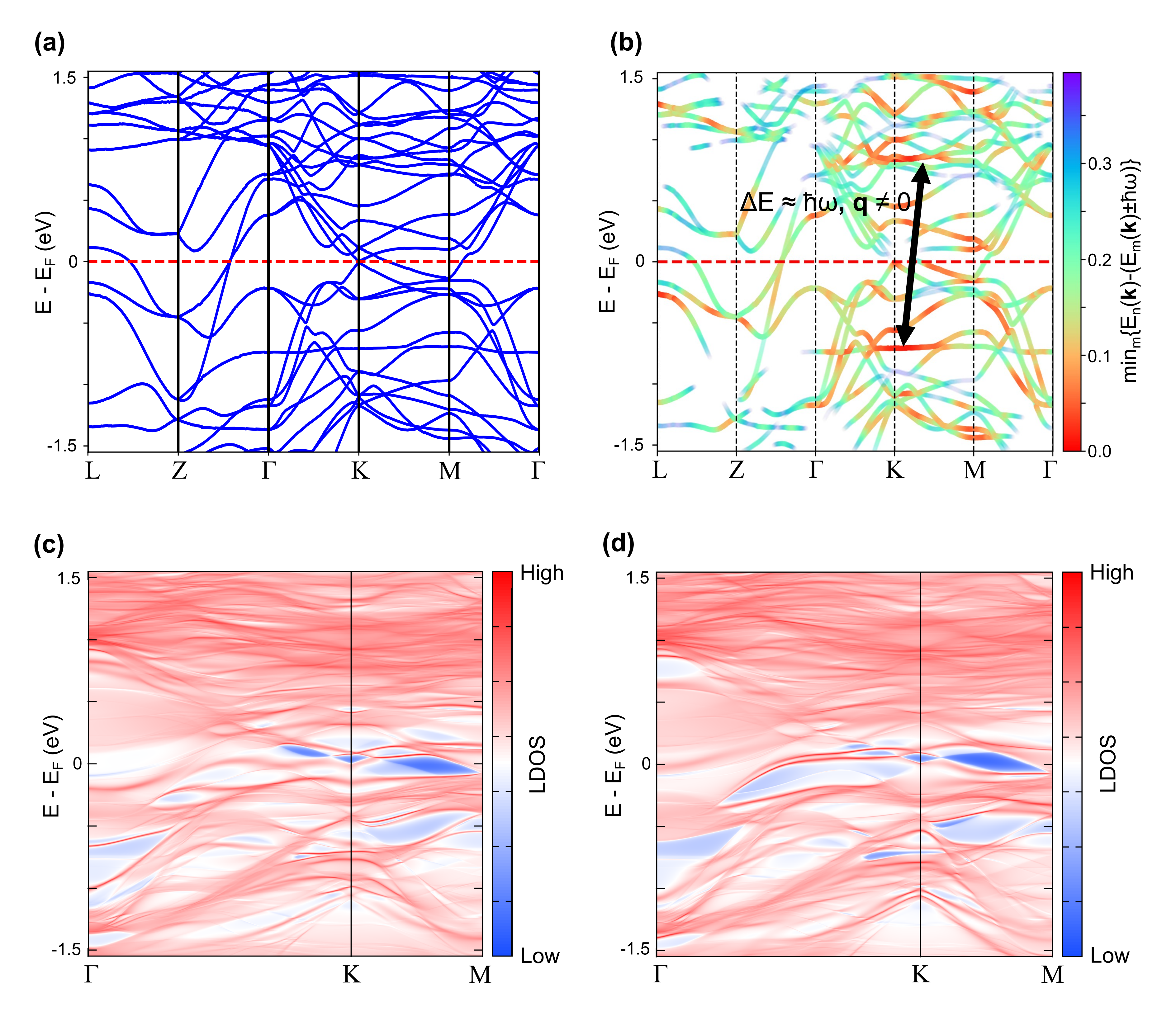}
    \caption{Bulk (\textbf{(a)} \& \textbf{(b)}) and surface (\textbf{(c)} \& \textbf{(d)}) band structure for Mn$_3$Sn. In \textbf{(a)} we show the bulk band structure, for an energy window corresponding to the frequency of the laser. In \textbf{(b)} we show the same plot, but with the color and opacity of the bands indicating the value of $\min_m \{E_n(\boldsymbol{k})-(E_m(\boldsymbol{k})\pm \hbar \omega) \}$ for band $n$, with $\pm$ indicating whether the band is above/below the Fermi surface. This is a rough indicator of the possibility of a vertical transition from band $n$ occurring. The flat bands around K may also allow non-vertical transitions. In \textbf{(c)} and \textbf{(d)} we show the surface band structure for the $c$-directed top and bottom surface respectively, with the color indicating how well-localised the states are on the surface.}
    \label{fig:fig4}
\end{figure}

%---------------------THEORY-----------------------------------
\clearpage
\newpage
\appendix
\section{Sample characterisation by X-ray diffraction analysis}

\begin{figure}[h]
    \centering
    \includegraphics[width=1\textwidth]{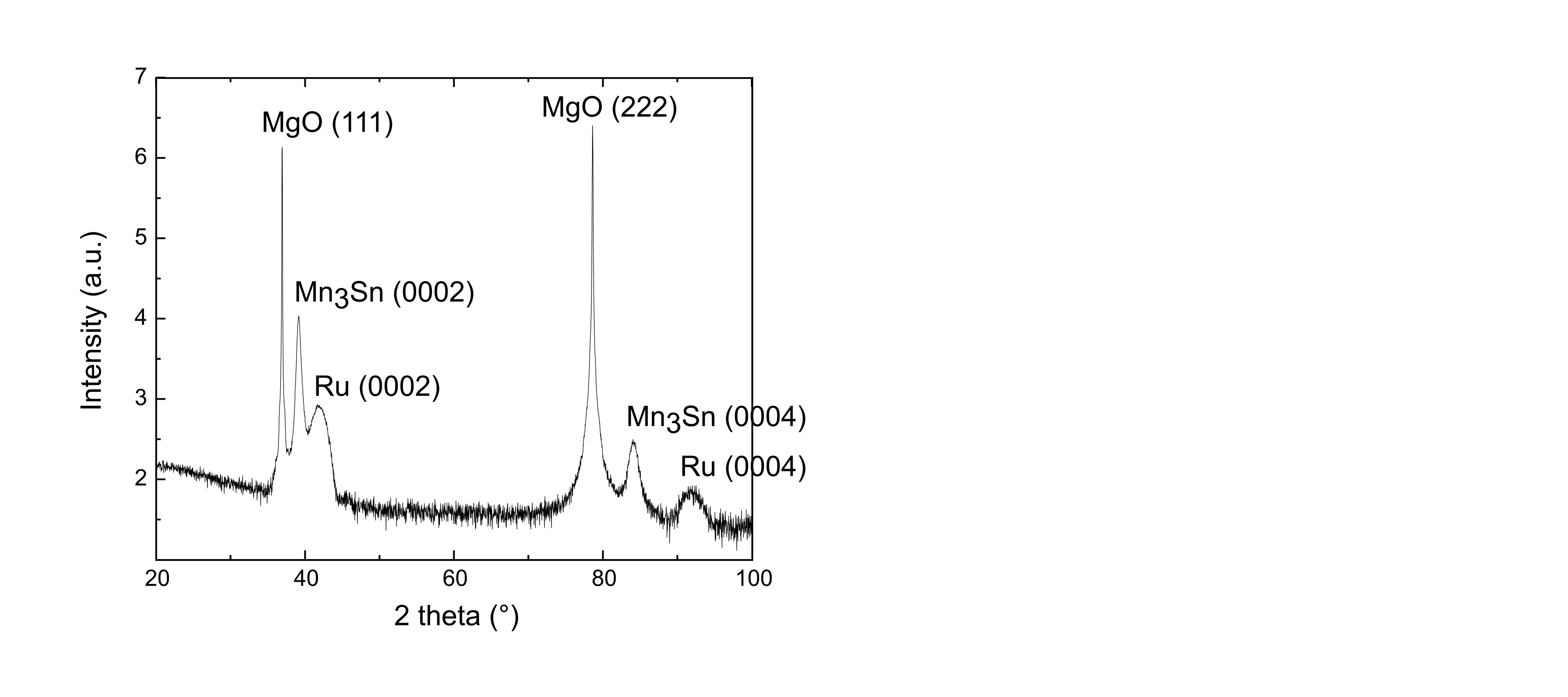}
    \caption
{
X-ray diffraction of the MgO(111)(0.5\,mm)/Ru(5\,nm)/Mn$_3$Sn(50\,nm)/Si(3\,nm) sample.
}
    \label{fig:figSXRD}
\end{figure}

\newpage
\section{Components of THz signal at normal incidence}

\begin{figure}[h]
    \centering
    \includegraphics[width=1\textwidth]{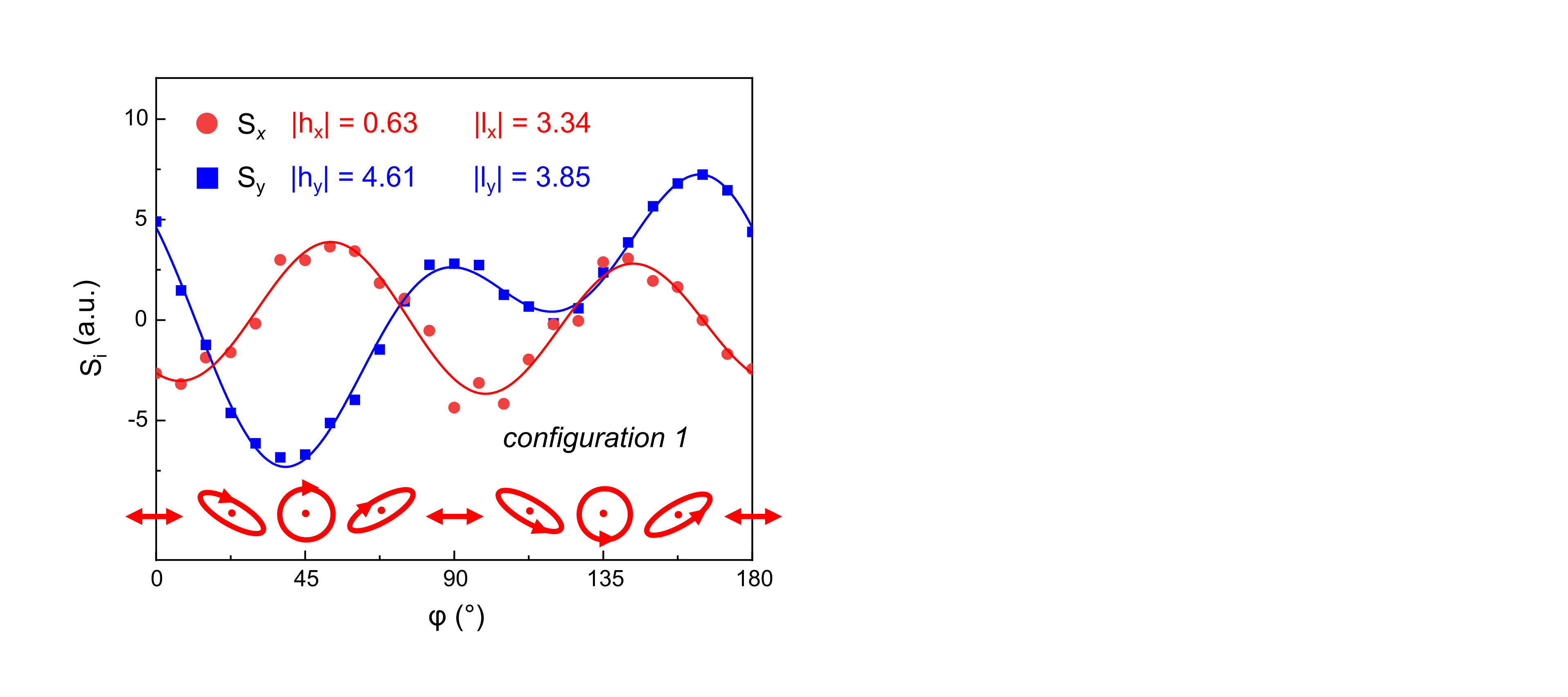}
    \caption
{
THz components emitted at normal laser incidence. THz signals measured for two orthogonal THz polarisations (along $x$ and along $y$) are plotted as a function of $\varphi$. The data sets were fitted with Eq.~(1) in the main text to extract the $h_i$ and $l_i$ parameters.
}
\label{fig:figS1}
\end{figure}

\newpage

\section{Even and odd THz responses measured along $x$}

\begin{figure}[h]
    \centering
    \includegraphics[width=1\textwidth]{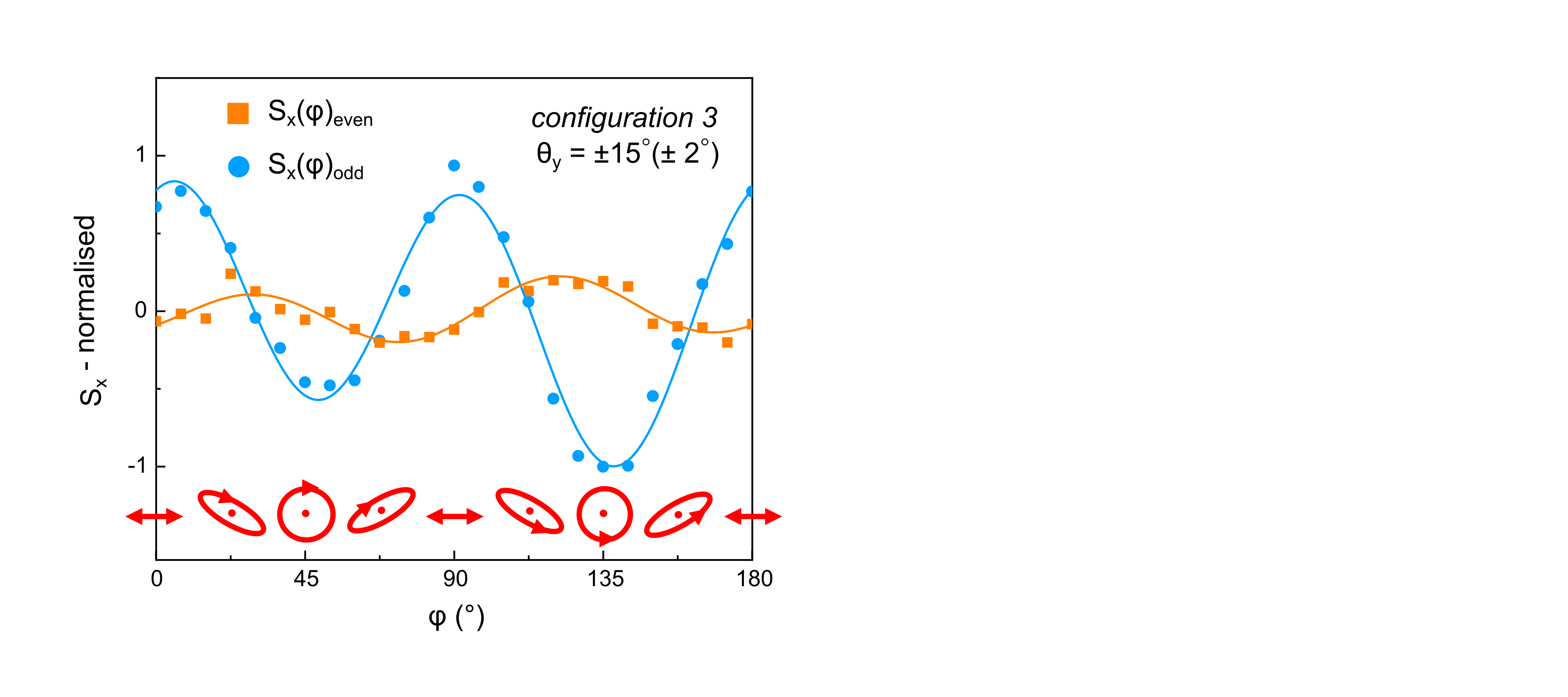}
    \caption
{
Even and odd THz responses measured along $x$ as a function of $\varphi$. The responses were obtained using Eq.~(2) and Eq.~(3) in the main text respectively. The experiment was performed in \textit{configuration 3}. 
}
    \label{fig:figS3}
\end{figure}

\newpage

\section{THz transients measured at different magnetic fields}

\begin{figure}[h]
    \centering
    \includegraphics[width=1\textwidth]{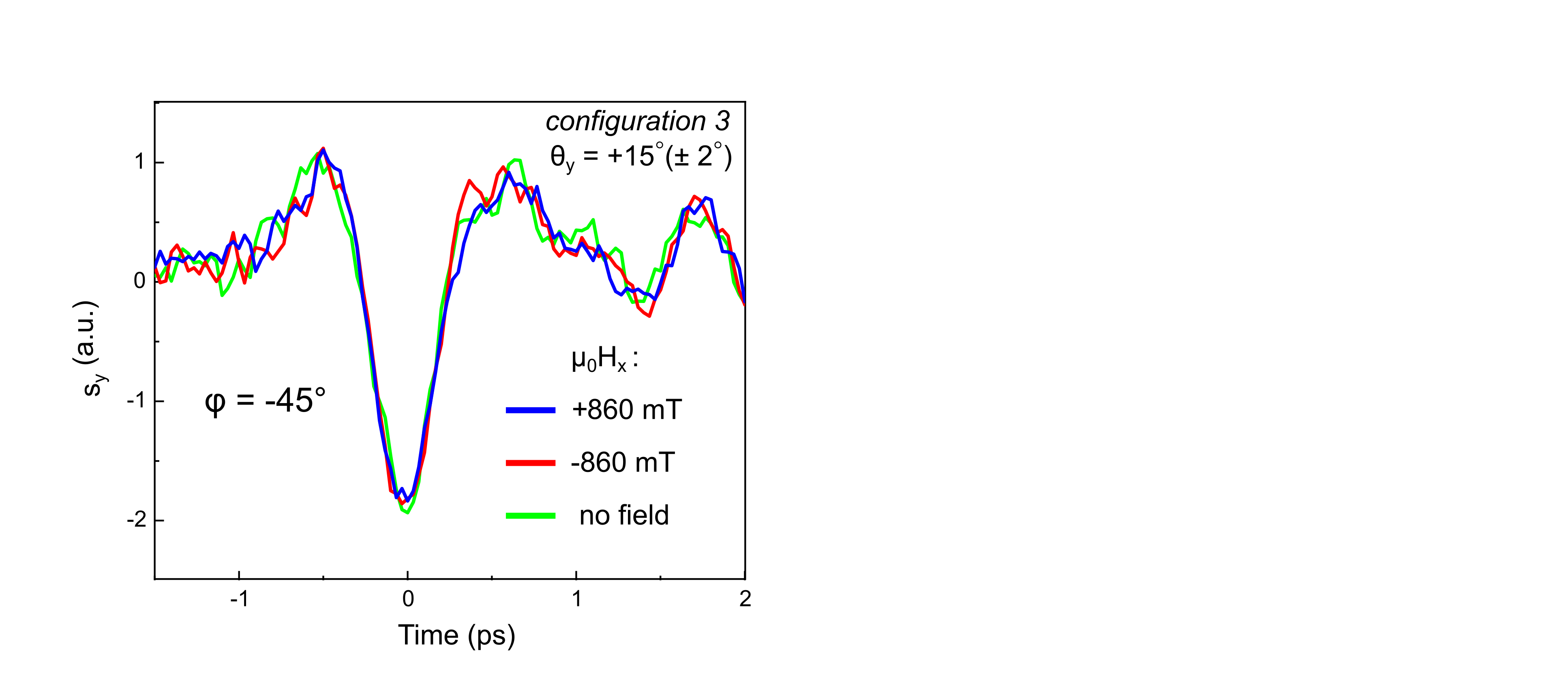}
    \caption
{
Transients THz fields detected for different directions of magnetic field and with no field applied. The measurement was performed at RT at a fixed position of the quarter wave plate ($\varphi=45\degree$), corresponding to RHCP.
}
    \label{fig:figS2}
\end{figure}

\newpage

\section{Temperature dependence of THz signal components}

\begin{figure}[h]
    \centering
    \includegraphics[width=1\textwidth]{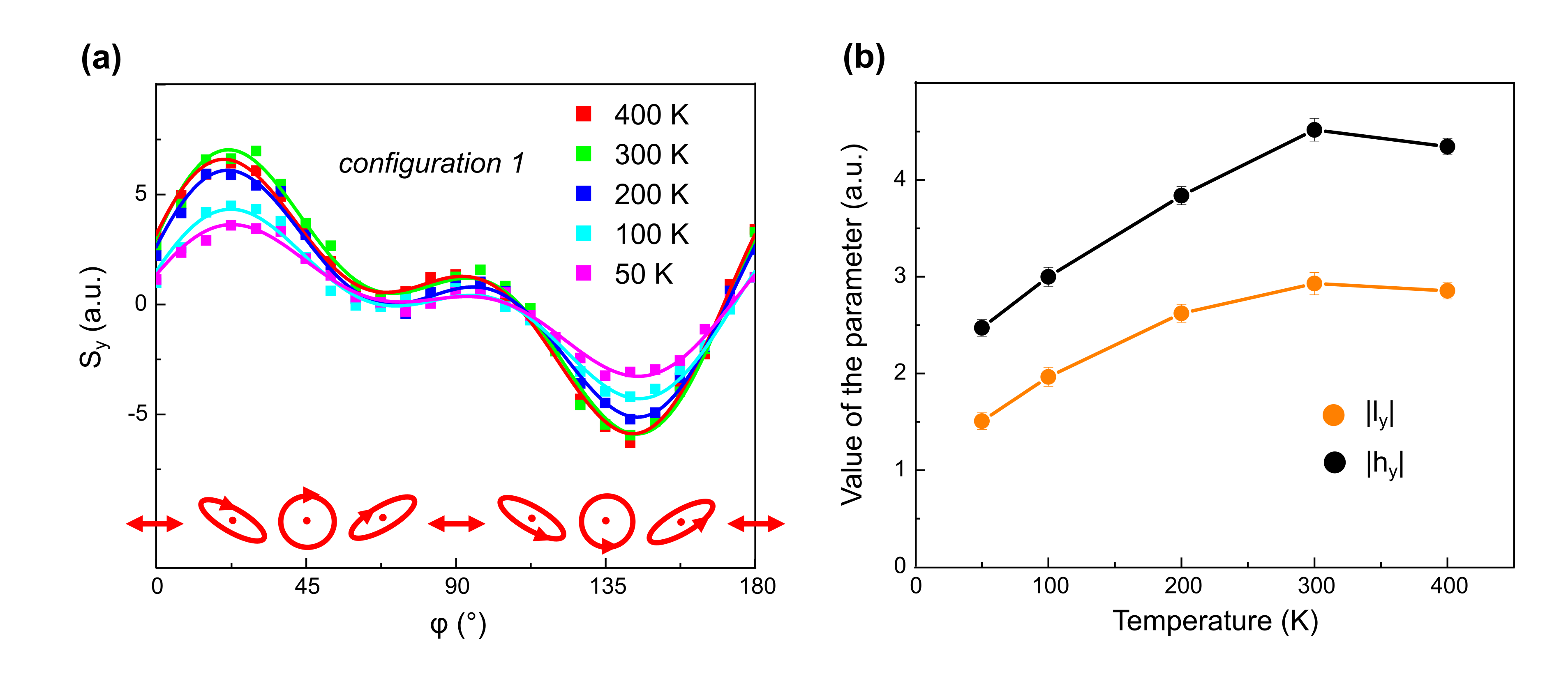}
    \caption
{
Temperature dependence of THz signal components. (a) shows peak THz signals measured along $y$ as a function of $\varphi$ at different temperatures between 50\,K and 400\,K. The measurement was performed in \textit{configuration 1} and with no magnetic field applied. The values of $|l_{y}|$ and $|h_{y}|$ were extracted from these data sets and are displayed in (b).
}
    \label{fig:figS4}
\end{figure}

\newpage

\section{Further theory details}
\subsection{Relevant space-groups}
Mn$_3$Sn crystallizes in a layered hexagonal lattice. When ignoring the magnetic moments, the bulk space group is the non-magnetic (gray) group P$6_3/$mmc$1'$ (No.194). Accounting for magnetism, the triangular antiferromagnetic phase stabilized between approximately 250\,K and 420\,K, is described by the magnetic space-group C$m'$c$m'$ (No. 63.464 in the BNS convention) albeit in a non-standard setting as shown in Ref.~\cite{Brown1990}.
\par
For the tensor symmetry analysis, only the point-groups (PG) matter. The non-magnetic bulk PG is 6/mmm$1'$, whereas the bulk magnetic PG is $m'm'm$. Because the photogalvanic effect (PGE) vanishes in the bulk, we also analyze the symmetry properties of the surface of the material. For all configurations, the light impinges on the surface perpendicular to the $c$-axis of the material. The non-magnetic PG associated with this surface is 3m$1'$, whereas the magnetic PG is simply $m'$. Note that this assumes that there is no reordering of the surface. We will argue that the observed signal arises predominantly from a bulk effect and as such this is not a major limitation. 

\subsection{Symmetry analysis of non-linear optical tensors}
This section discusses the full symmetry analysis for the optical response tensors considered in the main text. We begin by discussing the role of time-reversal symmetry, and then go on to discussing spatial symmetries. 

\subsubsection{Role of time-reversal symmetry}
The role of time-reversal in tensor symmetry analysis requires special care as described in Refs.~\cite{El-Batanouny2008,Eremenko1992,Gallego2019}. There are two distinct microscopic mechanisms (usually called "shift" and "injection" currents) \cite{Sipe2000} leading to induced currents in the PGE. These behave differently under time-reversal. As we are only interested in phenomenological expressions for the induced current, we do not consider the microscopic mechanisms. We therefore neglect antiunitary symmetries entirely when constraining our tensors, so that the analysis is agnostic to the underlying microscopic mechanism. The unitary PG in the non-magnetic case are 6/mmm in the bulk and 3m on the surface. In the magnetic case, the unitary part of the bulk PG is 2/m, whereas the unitary part of the surface PG is 1. 

\subsubsection{Role of spatial symmetries}
To understand how the remaining unitary point-group symmetries constrain the tensors $\chi_{ijk}^{(2)}$ and $\chi_{ijk}^{(3)}$ in Eq.~(4) of the main text, we first decompose the tensors describing the induced currents into symmetric and antisymmetric components with respect to the light field:
\begin{equation*}
J_i =  \chi^{(2),s}_{ijk}[E_j, E_k^*] + \chi^{(2),a}_{ik}P_{\mathrm{circ}}\hat{e}_k + \chi_{ijkl}^{(3),s}q_j[E_k,E_l^*] + \chi^{(3),a}_{ijk}q_jP_{\mathrm{circ}}\hat{e}_k
\end{equation*}
Where the symmetric (s) and antisymmetric (a) terms corresponds to the linear/circular PGE/PDE respectively. Following notation from Ref.~\cite{Karch2010} we have defined:
\begin{equation}
    [E_j,E_k^*] = \frac{E_jE_k^*+E_kE_j^*}{2} 
\end{equation}

\begin{equation}
    P_{\mathrm{circ}}\hat{e}_k = i(E\times E^*)_k
\end{equation}

The permutation symmetry of the tensors can be accounted for by using Jahn's symbols as discussed in \cite{Gallego2019}. The Jahn's symbol for each tensor is respectively $\chi_{i[jk]}^{(2),s}$: V[V2], $\chi_{ik}^{(2),a}$: V\{V2\}, $\chi_{ij[kl]}^{(3),s}$: V2[V2] and $\chi_{ijk}^{(3),a}$: V2\{V2\}. To find all symmetry-allowed terms, we use the \texttt{MTENSOR} tool \cite{Gallego2019} hosted on the \texttt{Bilbao Crystallographic Server (BCS}) \cite{Bilbao}. We write the resultant currents in the $(a,b,c)$ Cartesian coordinate system of the material. The full expressions for the induced current for all combinations of magnetic/non-magnetic, PDE/PGE and surface/bulk are in Sec.~\ref{sec:Full_expressions}.

\subsection{Dependence of signal on polarization and sample geometry}
\subsubsection{Dependence of incoming fields on angle and sample geometry}\label{sec:dependence_of_incoming}
We assume the laser is linearly polarized before passing through the quarter-wave plate, corresponding to a Jones vector \cite{Yariv2003} of $\hat{x}$. After passing through a quarter-wave plate at angle $\varphi$, we find in the laboratory frame:
\begin{equation}
    |E_x|^2 \propto \cos 4\varphi
\end{equation}
\begin{equation}
    |E_y|^2 \propto \cos 4\varphi
\end{equation}
\begin{equation}
    [E_x, E_y] \propto \sin 4\varphi
\end{equation}
\begin{equation}
    P_{circ}\hat{e}_k \propto [\sin 2\varphi]\hat{e}_k
\end{equation}
We assume that the THz emission follows the same polarization dependence, and therefore associate components in the measured signal with frequency $4\varphi$ with the LPGE/LPDE and components with frequency $2\varphi$ with the CPGE/CPDE. 

In the laboratory $(x,y,z)$ Cartesian frame, $\boldsymbol{E}=(E_x.E_y,0)$ and $\boldsymbol{q} = (0,0,1)$. Writing $\boldsymbol{\Tilde{E}},\boldsymbol{\Tilde{q}}$ for the quantities measured in the material $(a,b,c)$ Cartesian frame then gives for a rotation along the $x$-axis (\textit{configuration 2}):
\begin{equation}
    \boldsymbol{\Tilde{E}} = (E_x, E_y\cos \theta_x E_y, \sin \theta_x)
\end{equation}
\begin{equation}
    \boldsymbol{\Tilde{q}} = (0,q_z\sin \theta_x , q_z \cos \theta_x)
\end{equation}
\begin{equation}
    P_{\mathrm{circ}}\hat{e}_a = 0
\end{equation}
Whereas rotating along the $y$-axis (\textit{configuration 3}) gives:
\begin{equation}
    \boldsymbol{\Tilde{E}} = (E_x\cos \theta_y, E_y, E_x\sin \theta_y)
\end{equation}
\begin{equation}
    \boldsymbol{\Tilde{q}} = (q_z \sin \theta_y,0 , q_z \cos \theta_y)
\end{equation}
\begin{equation}
    P_{\mathrm{circ}}\hat{e}_b = 0
\end{equation}

\subsubsection{Expressions at normal incidence, $\theta_x=0,\theta_y=0$ (configuration 1)}
At normal incidence (\textit{configuration 1}), we can ignore optical corrections from refraction. From the full expression in Sec.~\ref{sec:Full_expressions}, we find for the induced bulk current when ignoring the magnetic ordering:
\begin{equation}
\begin{split}
J_a = 0\\
J_b = 0\\
J_c \propto q_z\cos 4 \varphi
\end{split}
\end{equation}
Whereas when we include the magnetic ordering, we find the bulk current:
\begin{equation}
    J_a \propto q_z \cos 4\varphi
\end{equation}
\begin{equation}
    J_b \propto  q_z(\sin 4\varphi+C_1\sin 2\varphi)
\end{equation}
\begin{equation}
    J_c \propto q_z \cos 4\varphi
\end{equation}
The allowed surface current at normal incidence when ignoring magnetic ordering are given by:
\begin{equation}
    J_a \propto \cos 4\varphi + C_2 q_z \cos 4\varphi
\end{equation}
\begin{equation}
    J_b \propto \sin 4\varphi + C_3 q_z \sin 4\varphi
\end{equation}
\begin{equation}
    J_c \propto \cos 4\varphi + C_4 q_z \cos 4\varphi
\end{equation}
When taking into account the magnetic ordering of the surface, all terms are symmetry-allowed. Here $C_i$ are constants that are independent of $\varphi$ and $\boldsymbol{q}$. We note that the non-magnetic analysis cannot explain the appearance of a circular effect at normal incidence, even when taking the surface into account. This is discussed further in the next section.

\subsubsection{Expressions away from  normal incidence (configuration 2 \& 3)}
As can be seen from the full expressions in Sec.~\ref{sec:Full_expressions}, away from normal incidence, all cases (magnetic/non-magnetic and surface/bulk) allow for a linear and a circular photocurrent, though all bulk contributions still arise exclusively from the PDE due to bulk inversion symmetry. Because the measured signal does not depend on the magnetic field (see Fig.~\ref{fig:fig3} and discussion in the main text as well as Fig.~\ref{fig:figS2}), we discuss only the non-magnetic symmetry settings in what follows.
\par
As the detector is in a fixed position, the detected signal will display a purely geometric variation under rotation away from normal incidence, as various faces of the crystal are exposed. The generated photocurrents will induce a dipole of strength $\boldsymbol{d}$, and the associated emitted THz field can be written in terms of unit vectors as:
\begin{equation}
    \boldsymbol{S}(\boldsymbol{r})\propto  \boldsymbol{E}^{\mathrm{THz}}(\boldsymbol{r)} \propto \hat{\boldsymbol{r}}\times \hat{\boldsymbol{d}}\times \hat{\boldsymbol{r}}
\end{equation}
Note that this ignores reflection and refraction effects, which will play a role away from normal incidence. As the sample is rotated by the same angles in both $x$ and $y$, however, these effects should be irrelevant when comparing \textit{configuration 2} and \textit{configuration 3}. The THz field is measured along the $z$-axis. The measured signal in \textit{configuration~1} is then:
\begin{equation}
   S_x \propto J_a
\end{equation}
\begin{equation}
    S_y \propto J_b
\end{equation}
Whereas in \textit{configuration 2} it is :
\begin{equation}
  S_x \propto J_a
\end{equation}
\begin{equation}
    S_y \propto J_b \cos \theta_x+J_c\sin \theta_x
\end{equation}
An in \textit{configuration 3}:
\begin{equation}
    S_x \propto J_a \cos \theta_y+J_c\sin \theta_y
\end{equation}
\begin{equation}
   S_y \propto J_b
\end{equation}
Where the induced currents $\boldsymbol{J}$ also depend on the angles $\theta_{x,y}$. This dependence can be written out explicitly, using the expressions in Sec.~\ref{sec:dependence_of_incoming} and Sec.~\ref{sec:Full_expressions}. This was shown for \textit{configuration 1} in all symmetry settings in the previous section. For the non-magnetic bulk signal in \textit{configuration 2}, we find:
\begin{equation}
    S_x \propto \sin \theta_x \cos \theta_x (C_1\sin 4\varphi +C_2 \sin 2\varphi)
\end{equation}
\begin{equation}
    S_y \propto (C_3 \cos \theta_x \sin \theta_x+C_4 \cos^3 \theta_x \sin \theta_x +C_5 \cos\theta_x \sin^3 \theta_x) \cos 4\varphi +(C_6\cos^3 \theta_x \sin \theta_x + C_7 \cos \theta_x \sin^3 \theta_x)\sin 4\varphi
\end{equation}
And in \textit{configuration 3}:
\begin{equation}
    S_x \propto (D_1\cos \theta_y \sin \theta_y + D_2\cos^3 \theta_y \sin \theta_y +D_3\cos \theta_y \sin^3 \theta_y) \cos 4\varphi +(D_4\cos^3 \theta_y \sin  \theta_y +D_5 \cos \theta_y \sin^3\theta_y)\sin 4\varphi
\end{equation}
\begin{equation}
S_y \propto \sin \theta_y \cos \theta_y (D_6\sin 4\varphi +D_7 \sin 2\varphi)
\end{equation}
Turning to the non-magnetic surface effects, the lowest order surface effects arises from the photogalvanic effect. For the surface photogalvanic effect, the angular dependence in \textit{configuration 2} is given by:
\begin{equation}
    S_x \propto (C_1^s+C_2^s\cos^2 \theta_x) \cos    4\varphi +C_3^s\sin \theta_x \sin 4\varphi +C_4^s\sin \theta_x \sin 2\varphi 
\end{equation}
\begin{equation}
    S_y \propto (C_5^s\sin \theta_x +C_6^s\cos^2 \theta_x \sin \theta_x +C_7^s\sin^3 \theta_x) \cos 4\varphi +(C_8^s \cos^2 \theta_x +C_9^s\cos^2 \theta_x \sin \theta_x)\sin 4\varphi
\end{equation}
And in \textit{configuration 3}:
\begin{equation}
    S_x\propto (D_1^s\cos \theta_y +D_2^s \sin \theta_y+ D_3^s\cos^3 \theta_y +D_4^s \sin^3 \theta_y +D_5^s \cos^2 \theta_y \sin \theta_y )\cos 4\varphi +D_6^s\cos^2 \theta_y \sin \theta_y \sin 4\varphi 
    \end{equation}
\begin{equation}
    S_y \propto (D_7^s \sin \theta_y +D_8^s\cos \theta_y)\cos 4\varphi +D_9^s\sin \theta_y \sin 2\varphi 
\end{equation}
In the above expressions, $C_i$ and $D_i$ ($C_i^s$ and $D_i^s$) are constant independent of $\theta_{x,y}$ and $\varphi$ associated with the generated bulk (surface) currents. 
\par
Note that there is no circular current being generated perpendicular to the rotation axis, independent of where the current is generated, which explains the large suppression in the circular current seen in Fig.~\ref{fig:fig2} of the main text. We attribute the appearance of a very small circular effect perpendicular to the rotation axis to an imperfect  sample alignment. All terms generated in the bulk are odd in the rotation angle, whereas some of the linear terms arising from the surface photogalvanic effect are even in rotation angle. By contrast, all circular terms are odd in rotation angle for both the surface and bulk effect. As shown in Fig.~\ref{fig:fig2} in the main text and Fig.~\ref{fig:figS3}, we see that the odd response dominates, even though the linear and circular components have comparable magnitudes. This suggests that the signal originates predominantly from the bulk photon drag effect.
\par
At normal incidence, we would expect to not detect any signal from the bulk photon drag effect. As shown in Fig.~\ref{fig:fig2}(c) of the main text, the detected signal at normal incidence is very weak compared to the signal at larger rotation angles. This suggests that the measured signal arises from an imperfect sample alignment, resulting in a small non-zero rotation angle. This also explains the observed signal in Fig.~\ref{fig:figS1}.

\subsubsection{Effect of in-plane rotation, $\theta_c \neq 0 $}
Rotating by $90 \degree$ in-plane (along the $c$-axis) keeps all optical parameters the same, so that all changes in signal arise as a result of the photocurrent generation mechanisms in the material. As shown in Fig.~\ref{fig:fig3}(b) in the main text, an in-plane rotation by $90\degree$ has negligible impact on the measured signal. Writing $R_{xa}$ for the rotation matrix relating the lab $(x,y,z)$ coordinates to the material $(a,b,c)$ coordinates, the measured signals are given by:
\begin{equation}
    \begin{split}
        S_x \propto \sum_{i \in (a,b,c)}R^{-1}_{xi}J_i(R\bold{E}, R\boldsymbol{q})\\
         S_y \propto \sum_{i \in (a,b,c)}R^{-1}_{yi}J_i(R\bold{E}, R\boldsymbol{q})
    \end{split}
\end{equation}
Inserting this into the equations found in Sec.~\ref{sec:Full_expressions}, we find that the non-magnetic bulk PDE is invariant under this transformation (in both the linear and circular component), whereas none of the other contributions are invariant under this transformation. This adds further credence to the result that our signal arises predominantly from a non-magnetic bulk PDE.

\section{Full expression for non-linear optical tensors}\label{sec:Full_expressions}
Here we provide full expressions for the symmetry-allowed currents, written in the material $(a,b,c)$ Cartesian coordinate system. These are found using the \texttt{MTENSOR} functionality \cite{Gallego2019} on the \texttt{BCS} \cite{Bilbao}.
\subsection{Non-magnetic bulk}
\subsubsection{Linear photon drag effect}
\begin{equation}
\begin{split}
J_a = q_a(\chi^{(3),s}_{aaaa}|E_a|^2 +\chi^{(3),s}_{aabb}|E_b|^2 + \chi^{(3),s}_{aacc}|E_c|^2)+\\
\frac{q_b}{2}\big( \chi^{(3),s}_{aaaa}-\chi^{(3),s}_{aabb}\big)[E_a, E_b]+q_c\chi^{(3),s}_{acac}[E_a, E_c]
\end{split}
\end{equation}
\begin{equation}
\begin{split}
J_b =  q_b(\chi^{(3),s}_{aabb}|E_a|^2 +\chi^{(3),s}_{aaaa}|E_b|^2 + \chi^{(3),s}_{aacc}|E_c|^2)+\\
    \frac{q_a}{2}\big( \chi^{(3),s}_{aaaa}-\chi^{(3),s}_{aabb}\big)[E_a, E_b]+q_c\chi^{(3),s}_{acac}[E_b, E_c]
\end{split}
\end{equation}
\begin{equation}
    \begin{split}
J_c =  q_c(\chi^{(3),s}_{ccaa}|E_a|^2 +\chi^{(3),s}_{ccaa}|E_b|^2 + \chi^{(3),s}_{cccc}|E_c|^2)+\\
    q_a \chi^{(3),s}_{caac} [E_a, E_c]+q_b \chi^{(3),s}_{caac} [E_b, E_c]
    \end{split}
\end{equation}
For a total of $7$ independent components.
\subsubsection{Circular photon drag effect}
\begin{equation}
        J_a = q_b\chi^{(3),a}_{abc}P_{\mathrm{circ}}\hat{e}_c-q_c\chi^{(3),a}_{acb}P_{\mathrm{circ}}\hat{e}_b
\end{equation}
\begin{equation}
    J_b = q_c\chi^{(3),a}_{acb}P_{\mathrm{circ}}\hat{e}_a-q_a\chi^{(3),a}_{abc}P_{\mathrm{circ}}\hat{e}_c
\end{equation}
\begin{equation}
    J_c = q_b\chi^{(3),a}_{cab}P_{\mathrm{circ}}\hat{e}_a-q_a\chi^{(3),a}_{cab}P_{\mathrm{circ}}\hat{e}_b
\end{equation}
For a total of 3 independent components
\subsection{Magnetic bulk}
\subsubsection{Linear photon drag effect}
\begin{equation}
\begin{split}
J_a = q_a(\chi^{(3),s}_{aaaa}|E_a|^2 +\chi^{(3),s}_{aabb}|E_b|^2 + \chi^{(3),s}_{aacc}|E_c|^2)+\\
q_a\chi^{(3),s}_{aaac}[E_a, E_c]+q_b\big(\chi^{(3),s}_{abbc}[E_b, E_c]+\chi^{(3),s}_{abab}[E_a, E_b]\big)+\\
 q_c(\chi^{(3),s}_{acaa}|E_a|^2 +\chi^{(3),s}_{acbb}|E_b|^2 + \chi^{(3),s}_{accc}|E_c|^2+\chi^{(3),s}_{acac}[E_a, E_c])
\end{split}
\end{equation}

\begin{equation}
\begin{split}
 J_b = q_b(\chi^{(3),s}_{bbaa}|E_a|^2 +\chi^{(3),s}_{bbbb}|E_b|^2 + \chi^{(3),s}_{bbcc}|E_c|^2)+\\
q_b\chi^{(3),s}_{bbac}[E_a, E_c]+q_a\big(\chi^{(3),s}_{babc}[E_b, E_c]+\chi^{(3),s}_{baab}[E_a, E_b]\big)+\\
 q_c(\chi^{(3),s}_{bcbc}[E_b,E_c]+\chi^{(3),s}_{bcab}[E_a, E_b])       
\end{split}
\end{equation}

\begin{equation}
\begin{split}
 J_c = q_a(\chi^{(3),s}_{caaa}|E_a|^2 +\chi^{(3),s}_{cabb}|E_b|^2 + \chi^{(3),s}_{cacc}|E_c|^2)+\\
q_a\chi^{(3),s}_{caac}[E_a, E_c]+q_b\big(\chi^{(3),s}_{cbbc}[E_b, E_c]+\chi^{(3),s}_{cbab}[E_a, E_b]\big)+\\
q_c(\chi^{(3),s}_{ccaa}|E_a|^2 +\chi^{(3),s}_{ccbb}|E_b|^2 + \chi^{(3),s}_{cccc}|E_c|^2+\chi^{(3),s}_{ccac}[E_a, E_c])  
\end{split}
\end{equation}
For a total of 28 independent components.

\subsubsection{Circular photon drag effect}
\begin{equation}
    \begin{split}
        J_a = -q_a\chi^{(3),a}_{aab}P_{\mathrm{circ}}\hat{e}_b-q_c\chi^{(3),a}_{acb}P_{\mathrm{circ}}\hat{e}_b\\
    +q_b(\chi^{(3),a}_{abc}P_{\mathrm{circ}}\hat{e}_c+\chi^{(3),a}_{aba}P_{\mathrm{circ}}\hat{e}_a)
    \end{split}
\end{equation}

\begin{equation}
    \begin{split}
        J_b = q_a(\chi^{(3),a}_{bac}P_{\mathrm{circ}}\hat{e}_c+\chi^{(3),a}_{baa}P_{\mathrm{circ}}\hat{e}_a)\\
        +q_c(\chi^{(3),a}_{bcc}P_{\mathrm{circ}}\hat{e}_c+\chi^{(3),a}_{bca}P_{\mathrm{circ}}\hat{e}_a)\\
        -q_b\chi^{(3),a}_{bbb}P_{\mathrm{circ}}\hat{e}_b
    \end{split}
\end{equation}

\begin{equation}
\begin{split}
            J_c = -q_a\chi^{(3),a}_{cab}P_{\mathrm{circ}}\hat{e}_b-q_c\chi^{(3),a}_{ccb}P_{\mathrm{circ}}\hat{e}_b\\
        +q_b(\chi^{(3),a}_{cbc}P_{\mathrm{circ}}\hat{e}_c+\chi^{(3),a}_{cba}P_{\mathrm{circ}}\hat{e}_a)
\end{split}
\end{equation}
For a total of 13 independent coefficients.

\subsection{Non-magnetic surface}
\subsubsection{Linear photogalvanic effect}
\begin{equation}
    J_a =  \chi^{(2),s}_{aaa}(|E_a|^2-|E_b|^2)+\chi^{(2),s}_{aac}[E_a, E_c]
\end{equation}
\begin{equation}
    J_b = \chi^{(2),s}_{aac}[E_b, E_c]-\chi^{(2),s}_{aaa}[E_a, E_b]
\end{equation}
\begin{equation}
    J_c = \chi^{(2),s}_{caa}(|E_a|^2+|E_b|^2)+\chi^{(2),s}_{ccc}|E_c|^2
\end{equation}
For a total of 4 independent coefficients.
\subsubsection{Circular photogalvanic effect}
\begin{equation}
    J_a = -\chi^{(2),a}_{ab}P_{\mathrm{circ}}\hat{e}_b
\end{equation}
\begin{equation}
    J_b = \chi^{(2),a}_{ab}P_{\mathrm{circ}}\hat{e}_a
\end{equation}
\begin{equation}
    J_c = 0
\end{equation}
With a single independent parameter.
\subsubsection{Linear photon drag effect}
\begin{equation}
\begin{split}
J_a =  q_a(\chi^{(3),s}_{aaaa} |E_a|^2 +\chi^{(3),s}_{aabb} |E_b|^2 + \chi^{(3),s}_{aacc} |E_c|^2)\\
+q_a\chi^{(3),s}_{aaac}[E_a, E_c]+\frac{q_b}{2}\big(\chi^{(3),s}_{aaaa}-\chi^{(3),s}_{aabb} \big)[E_a,E_b]\\
-q_b\chi^{(3),s}_{aaac}[E_b, E_c]+q_c \chi^{(3),s}_{acaa}\big(|E_a|^2-|E_b|^2\big) + q_c\chi^{(3),s}_{acac}[E_a, E_c]
\end{split}
\end{equation}

\begin{equation}
\begin{split}
     J_b = q_b(\chi^{(3),s}_{aabb} |E_a|^2 + \chi^{(3),s}_{aaaa}|E_b|^2 + \chi^{(3),s}_{aacc} |E_c|^2)\\
     -q_b\chi^{(3),s}_{aaac}[E_a, E_c]-q_a\chi^{(3),s}_{aaac}[E_b,E_c]\\
     +\frac{q_a}{2}\big(\chi^{(3),s}_{aaaa}-\chi^{(3),s}_{aabb})[E_a,E_b]+q_c(\chi^{(3),s}_{acac}[E_b,E_c]-\chi^{(3),s}_{acaa}[E_a,E_b])
\end{split}
\end{equation}

\begin{equation}
\begin{split}
        J_c = q_c(\chi^{(3),s}_{ccaa} (|E_a|^2+|E_b|^2)+\chi^{(3),s}_{cccc}|E_c|^2)\\
    +q_a\big(\chi^{(3),s}_{caaa}(|E_a|^2-|E_b|^2)+\chi^{(3),s}_{caac}[E_a,E_c]\big)\\
    +q_b(\chi^{(3),s}_{caac}[E_b,E_c]-\chi^{(3),s}_{caaa}[E_a,E_b])
\end{split}
\end{equation}
For a total of 10 independent components. 
\subsubsection{Circular photon drag effect}
\begin{equation}
\begin{split}
        J_a = q_bP_{\mathrm{circ}}(\chi^{(3),a}_{abc}\hat{e}_c-\chi^{(3),a}_{aab}\hat{e}_a)\\ 
        -\chi^{(3),a}_{aab}q_aP_{\mathrm{circ}}\hat{e}_b-\chi^{(3),a}_{acb}q_cP_{\mathrm{circ}}\hat{e}_b
        \end{split}
\end{equation}

\begin{equation}
\begin{split}
J_b = -q_aP_{\mathrm{circ}}(\chi^{(3),a}_{abc}\hat{e}_c+\chi^{(3),a}_{aab}\hat{e}_a)\\
+q_bP_{\mathrm{circ}} \chi^{(3),a}_{aab}\hat{e}_b+q_cP_{\mathrm{circ}} \chi^{(3),a}_{acb}\hat{e}_a
\end{split}
\end{equation}

\begin{equation}
    J_c = \chi^{(3),a}_{cab} P_{\mathrm{circ}}(q_b\hat{e}_a-q_a\hat{e}_b)
\end{equation}
For a total of 4 independent components.
\subsection{Magnetic surface}
As the unitary part of the symmetry group is $1$ on the surface when considering magnetic symmetries, all terms are allowed. 

\par
\bibliography{references}
\end{document}